\documentclass[aip,jap,reprint,graphicx]{revtex4-1}
\usepackage{color,graphicx}
\usepackage{rcs}
\usepackage{color,graphicx}
\usepackage[dvipsnames]{xcolor}
\usepackage{dcolumn}
\usepackage{bm}
\usepackage{chngcntr}
\usepackage[thinspace,thinqspace]{SIunits}
\draft 

\begin{document}


\title{Modes of an elliptical cylindrical resonant cavity - analytical solution} 



\author{M. S. Grbi\'{c}}
\email[]{mgrbic@phy.hr}
\affiliation{Department of Physics, Faculty of Science, University
	of Zagreb, Bijeni\v {c}ka 32, Zagreb HR 10000, Croatia}


\date{\today}

\begin{abstract}
An analytical solution of the Helmholtz equation for electromagnetic field distribution in a resonant cavity with elliptic cross-section is found. We compare the frequencies of the eigenmodes with  numerical and experimental values for a metallic cavity and find an excellent matching. We focus our analysis on the microwave frequency region, and show how the ellipticity of the cavity (ratio of the minor and major axes length $b/a$) influences several mode frequencies and also the $Q$-factor of the cavity. By doing so, we demonstrate how the elliptic geometry splits the degeneracy of certain modes of the circular cylindric cavity.
\end{abstract}


\maketitle 

\section{INTRODUCTION}
Resonant cavities have been widely used to study the macroscopic and microscopic electromagnetic properties of materials~\cite{Poole,Nebendahl}, acoustic modes of gas pulsations~\cite{Hong}, optics~\cite{Tuan} and quantum billiards~\cite{Tuan,Waalkens}, but also in the design of particle accelerators~\cite{TESLAcavity}. Complete understanding of the eigenmodes for a specific geometry is therefore important for design of future systems, but also for proper data analysis. In condensed matter physics, for the practical purposes and restrictions on sample space in the cryostats, the geometry of the cavity is often chosen to be simple - rectangular or circular cylinder. However, this simplification in some cases creates limitations, e.g. in the case of the rectangular cavity, the available sample space inside the cavity is limited by the outer dimension set by the bore size inside the cryostat. Another example of limitations is the mode degeneracy of the circular cylinder that obscures the true distribution of the electromagnetic fields inside the cavity. This can make data analysis difficult if the cavity is used for studying modern systems.\\
\indent Here we present an analytical solution for the electromagnetic eigenmode frequencies and magnetic (electric) field distributions of an elliptical cylindrical cavity where the degeneracy of the circular case is broken. We show how the frequencies of specific modes depend on geometry, or more specific, eccentricity of the elliptical cross-section. In the paper we focus on the eigenmodes in the microwave frequency region and compare our results with numerical simulations and experimental data. Due to a higher level of mathematical complexity, existing analysis of this problem is sparse, and with no comparison of the results with experimental values. Initial analysis of a similar problem was done by Chu~\cite{Chu} albeit for waveguides. Kinzer and Wilson~\cite{Kinzer} worked on the elliptic cavity problem, but their solution is made suitable for an approximation by series expansion of Bessel functions, which is then used for calculation of corresponding numerical values. Higgins and Straiton~\cite{Higgins} have analyzed only a specific (transverse electric) mode, while we provide a complete solution for all modes.

The broken degeneracy of the elliptical cavity has proven to be useful in studying the properties of cuprate superconductors~\cite{Peligrad1998,Peligrad2001,Grbic09,Grbic11} with large anisotropy of the sample's conductivity that requires the exact knowledge of the electromagnetic field distribution to properly treat the measured data.


\section{GENERAL SOLUTION OF THE WAVE EQUATION IN ELLIPTIC COORDINATES}
\begin{figure}
\includegraphics[width=0.5\textwidth]{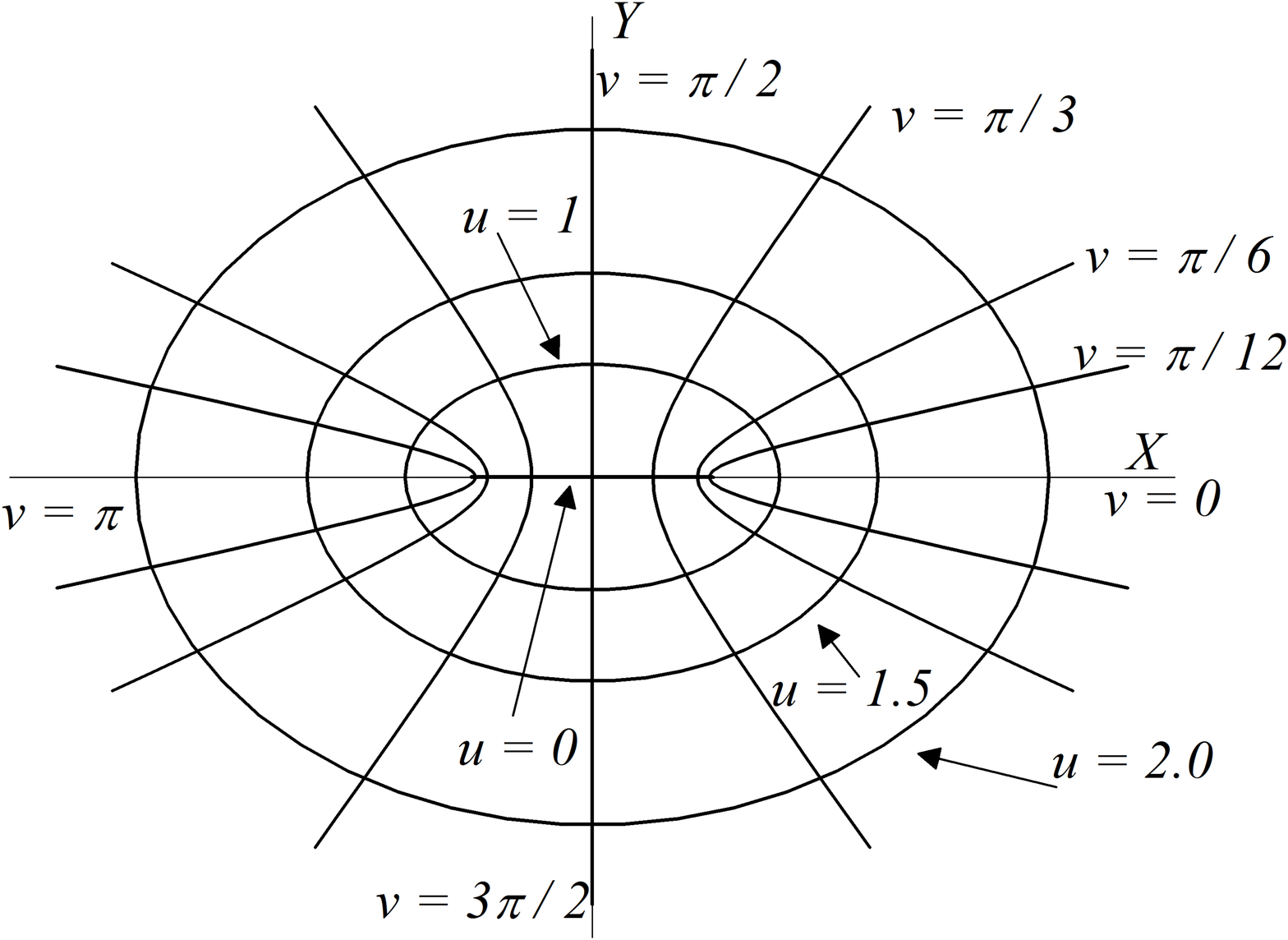}
 \caption{Elliptical coordinates. Variable $u$ is similar to the radial parameter $r$ in the polar coordinate system, and variable $v$ to the angular parameter $\varphi$.  }%
 \label{Fig1}
 \end{figure}

The linear homogeneous Helmholtz equation for a three-dimensional cavity 
can be written as:
\begin{equation}\label{eq:Helmh}
\nabla^2 {\bf{\Psi}}(x,y,z,t) + \mu_0 \varepsilon_0 \omega^2 {\bf{\Psi}}(x,y,z,t) = 0,
\end{equation}
where $\omega$ is the frequency, $\mu_0$ magnetic permeability and $\varepsilon_0$ electric permittivity. $\bf{\Psi}$ marks the solution in the form of electric or magnetic fields. The eigenmodes here can be divided into transverse-electric (TE) and transverse-magnetic (TM) modes. TE modes have no $z$ component of the electric field ($\bf{E}$), while TM modes have no $z$ component of the magnetic field ($\bf{H}$). Because of the symmetry of the problem, we need to use the elliptical cylindrical coordinates, presented in Fig.~\ref{Fig1}. The coordinates can be related to the Cartesian system by the equations~\cite{McLachlan}:
\begin{eqnarray}
&&x=f \cosh (u) \cos (v), \\
&&y=f \sinh (u) \sin (v), \\
&&z=z,
\end{eqnarray}
where $f$ is the semi-interfocal distance of the ellipse, while $u$ and $v$ are the elliptic coordinates. If we define by $a$ and $b$ as the half-lengths of the major and minor axes, respectively, then the eccentricity of the ellipse bounded by $u=u_0$ is $e=f/a=1/\cosh(u_0)=\sqrt{1-(b/a)^2}$.

Let us consider that our elliptical cylindrical cavity has boundaries located at $u=u_0$, $z=0$ and $L$, and that its walls are made of a rigid highly-conductive medium. By assuming that the time dependence of the fields can be separated in the form ${\bf{\Psi}}(u,v,z,t)={\bf{\Psi}}_s (u,v,z) e^{i(\omega t+ \varphi)}$ we can re-write Eq.~(\ref{eq:Helmh}) in the elliptical coordinates for the $z$ component of the fields, and then derive the others from Maxwell equations~\cite{McLachlan}. With this in mind, we write for the TE modes: 
\begin{eqnarray}\label{eq:HelmhEll}
\frac{2}{f^2 (\cosh(2 \textit{u})-\cos(2 \textit{v}))} && \left(\frac{\partial^2 H_z}{\partial u^2} +\frac{\partial^2 H_z}{\partial v^2}\right) +\frac{\partial^2 H_z}{\partial z^2} + \nonumber \\  &&+ \mu_0 \varepsilon_0 \omega_{TE} ^2 H_z = 0,
\end{eqnarray}
and similar for the TM modes
\begin{eqnarray}
\frac{2}{f^2 (\cosh(2 \textit{u})-\cos(2 \textit{v}))} &&\left(\frac{\partial^2 E_z}{\partial u^2}+\frac{\partial^2 E_z}{\partial v^2}\right)+\frac{\partial^2 E_z}{\partial z^2} + \nonumber \\
&&+ \mu_0 \varepsilon_0 \omega_{TM} ^2 E_z = 0.
\end{eqnarray}
By using separation of variables, we can write:
\begin{eqnarray}
H_z ^{TE}(u,v,z) =H_z ^{TE} (u,v) Z ^{TE}(z),\\
E_z ^{TM}(u,v,z) =E_z ^{TM} (u,v) Z ^{TM}(z)
\end{eqnarray}
for the TE and TM modes, respectively. The above expressions can then be written in the form:
\begin{eqnarray}
&&\frac{2}{f^2 (\cosh(2 \textit{u})-\cos(2 \textit{v}))} \left(\frac{\partial^2 H_z ^{TE}}{\partial u^2} +\frac{\partial^2 H_z ^{TE}}{\partial v^2}\right) \frac{1}{H_z ^{TE}} +\nonumber \\  &&+\frac{1}{Z^{TE}(z)}\frac{\partial^2 Z^{TE}(z)}{\partial z^2} + \mu_0 \varepsilon_0 \omega_{TE} ^2 = 0,
\end{eqnarray}
and similar for the TM modes
\begin{eqnarray}
&&\frac{2}{f^2 (\cosh(2 \textit{u})-\cos(2 \textit{v}))} \left(\frac{\partial^2 E_z ^{TM}}{\partial u^2}+\frac{\partial^2 E_z ^{TM}}{\partial v^2}\right) \frac{1}{E_z ^{TM}}+ \nonumber \\
&&+\frac{1}{Z^{TM}(z)}\frac{\partial^2 Z^{TM}(z)}{\partial z^2} + \mu_0 \varepsilon_0 \omega_{TM} ^2 = 0.
\end{eqnarray}
These can be separated in the same manner:
\begin{eqnarray}
&&\frac{2}{f^2 (\cosh(2 \textit{u})-\cos(2 \textit{v}))} \left(\frac{\partial^2 H_z ^{TE}}{\partial u^2} +\frac{\partial^2 H_z ^{TE}}{\partial v^2}\right) \frac{1}{H_z ^{TE}} -\nonumber \\ 
&&-k_{TE} ^2 + \mu_0 \varepsilon_0 \omega_{TE} ^2 = 0,
\end{eqnarray}
\begin{eqnarray}
\frac{\partial^2 Z^{TE}(z)}{\partial z^2} = - k_{TE} ^2 Z^{TE}(z),
\end{eqnarray}
and similar for the TM modes
\begin{eqnarray}
&&\frac{2}{f^2 (\cosh(2 \textit{u})-\cos(2 \textit{v}))} \left(\frac{\partial^2 E_z ^{TM}}{\partial u^2}+\frac{\partial^2 E_z ^{TM}}{\partial v^2}\right) \frac{1}{E_z ^{TM}}- \nonumber \\
&&-k_{TM} ^2 + \mu_0 \varepsilon_0 \omega_{TM} ^2 = 0,
\end{eqnarray}
\begin{eqnarray}
\frac{\partial^2 Z^{TM}(z)}{\partial z^2} = - k_{TM} ^2 Z^{TM}(z),
\end{eqnarray}
using $k_i$ ($i=$ TE, TM) as a separation constant. By applying the boundary conditions $Z ^{TE}(z=0)= Z ^{TE}(z=L)=0$, and $\left. \partial Z ^{TM}(z)/\partial z\right|_{z=0} = \left. \partial Z ^{TM}(z)/\partial z\right|_{z=L} =0 $, we find $k_{i}=\pi m_{i}/L$ for both $i=$TE, TM modes, where $m_i = 1,2,3...$ for TE modes and $m_i = 0,1,2,3...$ for TM modes. The solutions can then be written as
\begin{eqnarray}
H_z(u,v,z) =H_z ^{TE} (u,v) \sin(k_{TE} z),\\
E_z(u,v,z) =E_z ^{TM} (u,v) \cos(k_{TM} z).
\end{eqnarray}
The earlier expressions can now be written in the simpler form
\begin{eqnarray}
&&\frac{\partial^2 H^{TE}_z}{\partial u^2}+\frac{\partial^2 H^{TE}_z}{\partial v^2}+\nonumber \\
&&+\frac{1}{2} {k_{1}}^2 f^2 (\cosh(2 u)-\cos(2 v)) H^{TE}_z = 0,
\end{eqnarray}
with $k_1^2 = -k_{TE}^2 +\mu_0 \varepsilon_0 \omega_{TE} ^2$ and
\begin{eqnarray}
&&\frac{\partial^2 E^{TM}_z}{\partial u^2}+\frac{\partial^2 E^{TM}_z}{\partial v^2}+\nonumber \\
&&+\frac{1}{2} {\kappa_{1}}^2 f^2 (\cosh(2 u)-\cos(2 v)) E^{TM}_z = 0,
\end{eqnarray}
with $\kappa_1^2 = -k_{TM}^2 +\mu_0 \varepsilon_0 \omega_{TM} ^2$.
These equations can be further separated to reach a form of Mathieu equations (for variable $v$) and modified Mathieu equations (for variable $u$), which can be found in the Appendix~\ref{app:Separ}.  Now, for simplicity we will abandon the current procedure of following both TE and TM modes, and focus on the TE modes to reach the final form of the solution. After we establish the procedure, we will return to the TM modes and analyze the results.
\section{NATURAL MODES OF THE ELLIPTICAL CYLINDRICAL CAVITY}
\subsection{TE modes}
The general form of the solution for the $H_z ^{TE}(u,v)$ field of the TE modes can be written as
\begin{eqnarray}
H_z ^{TE}(u,v) = \sum^{\infty}_{n=0} C_n Ce_n (u,q) ce_n (v,q)+\nonumber \\ + \sum^{\infty}_{n=1} S_n Se_n (u,q) se_n (v,q),
\label{eq:HzTE}
\end{eqnarray}  
if we make the substitution $4q = k_1^2 f^2$. Here $C_n , S_n$ are constants; $n$ marks the mode index; $ce_n, se_n$ are the ordinary and $Ce_n, Se_n$ the modified Mathieu functions. The value of $q$ is determined by the boundary conditions, which is in the current case $\partial H_z /\partial u (u=u_0)= 0$ at the inner surface of the cavity. This leads to two conditions:
\begin{eqnarray}
&&\left. \partial Ce_n (u,q) /\partial u \right|_{u=u_0}= Ce_n ' (u=u_0,q) =0, \label{eq:boundTE}\\ 
&&\left.\partial Se_n(u,q) /\partial u \right|_{u=u_0} = Se_n ' (u=u_0,q) =0,\label{eq:boundTE2}
\end{eqnarray}
i.e. for TE modes we need to find the roots $q_{n,p}$ and $\overline{q}_{n,p}$ that give rise to zeros of the first derivative of the modified Mathieu functions $Ce_n$ and $Se_n$. $n,p$ mark the $p$-th root of the $n$-th Mathieu function.

All other components of \textbf{H}$^{TE} (u,v,z)$ (and \textbf{E}$^{TE}(u,v,z)$) can be calculated from $H_z ^{TE} (u,v,z)$ and the condition $E_z ^{TE} (u,v,z)=0$ from the Maxwell equations. We will now present the complete solution of the TE modes, albeit using a more compact form. With the summation across $n$ as it is done in~(\ref{eq:HzTE}) and $l(u,v) = f \sqrt{\cosh(2 u)-\cos(2v)}/\sqrt{2}$, we now write all the components of the eigenmode's magnetic field  as:
\begin{eqnarray}
&&H_z (u,v,z) =\nonumber \\
&&= \left\{
\begin{array}{ll}
C_{n,p} Ce_n(u,q_{n,p}) ce_n(v,q_{n,p}) \\
S_{n,p} Se_n(u,\bar{q}_{n,p}) se_n(v,\bar{q}_{n,p}) 
\end{array} 
\right\} \sin(k_{TE} z), 
\end{eqnarray}
\begin{eqnarray}
&&H_u (u,v,z) =\frac{k_{TE} }{k_{1}^2 l(u,v)} \times\\
&& \times \left\{
\begin{array}{c}
C_{n,p} Ce_n'(u,q_{n,p}) ce_n(v,q_{n,p}) \\
S_{n,p} Se_n'(u,\bar{q}_{n,p}) se_n(v,\bar{q}_{n,p}) \nonumber
\end{array} 
\right\} \cos(k_{TE} z) ,
\end{eqnarray}
\begin{eqnarray}
&&H_v (u,v,z) =\frac{k_{TE} }{k_{1}^2 l(u,v)} \times\\
&& \times \left\{
\begin{array}{c}
C_{n,p} Ce_n(u,q_{n,p}) ce'_n(v,q_{n,p}) \\
S_{n,p} Se_n(u,\bar{q}_{n,p}) se'_n(v,\bar{q}_{n,p}) \nonumber
\end{array} 
\right\} \cos(k_{TE} z) .
\end{eqnarray}
For the components of the electric field it follows:
\begin{eqnarray}
E_z (u,v,z) = 0
\end{eqnarray}
\begin{eqnarray}
&&E_u (u,v,z) =\frac{\mu_0 \omega }{k_{1}^2 l(u,v)} \times\\
&& \times \left\{
\begin{array}{c}
C_{n,p} Ce_n(u,q_{n,p}) ce'_n(v,q_{n,p}) \\
S_{n,p} Se_n(u,\bar{q}_{n,p}) se'_n(v,\bar{q}_{n,p}) \nonumber
\end{array} 
\right\} \sin(k_{TE} z) ,
\end{eqnarray}
\begin{eqnarray}
&&E_v (u,v,z) =-\frac{\mu_0 \omega }{k_{1}^2 l(u,v)} \times\\
&& \times \left\{
\begin{array}{c}
C_{n,p} Ce'_n(u,q_{n,p}) ce_n(v,q_{n,p}) \\
S_{n,p} Se'_n(u,\bar{q}_{n,p}) se_n(v,\bar{q}_{n,p}) \nonumber
\end{array} 
\right\} \sin(k_{TE} z) .
\end{eqnarray}
The frequencies of the TE modes can now be written in the form
\begin{eqnarray}
\omega_{n,p}^{TE}=\frac{1}{\mu_0 \varepsilon_0} \sqrt{\left(\frac{\pi m}{L}\right)^2 + 4\frac{q(n,p)}{f^2}},
\end{eqnarray}
where $q(n,p) = q_{n,p}$ and $\bar{q}_{n,p}$, for a specific mode. In the case of $q_{n,p}$, $p =1, 2, 3...$ and for $\bar{q}_{n,p}$, $p= 0, 1, 2, 3...$
\subsection{TM modes}
For the TM modes the general form of the solution for the $E_z ^{TM}(u,v)$ field can be written as
\begin{eqnarray}
E_z ^{TM}(u,v) = \sum^{\infty}_{n=0} D_n Ce_n (u,Q) ce_n (v,Q)+\nonumber \\ + \sum^{\infty}_{n=1} T_n Se_n (u,Q) se_n (v,Q),
\end{eqnarray}  
where $D_n , T_n$ are constants; $4Q = \kappa_1^2 f^2$, and the value of $Q$ is again determined by the boundary conditions, which is now $E_z~(u=u_0)=0$, at the inner surface. This leads to another two conditions:
\begin{eqnarray}
&&Ce_n  (u=u_0,Q) =0, \label{eq:boundTM}\\ 
&& Se_n (u=u_0,Q) =0. \label{eq:boundTM2}
\end{eqnarray}
Hence, for the TM modes we need to find the roots $Q_{n,p}$ and $\overline{Q}_{n,p}$ that give rise to zeros of the modified Mathieu functions $Ce_n$ and $Se_n$.

Using the same compact form, we can now show the complete solution of the TM modes. With previous definitions we can write for the components of the electric field:
\begin{eqnarray}
&&E_z (u,v,z) =\\
&&= \left\{
\begin{array}{c}
D_{n,p} Ce_n(u,Q_{n,p}) ce_n(v,Q_{n,p}) \\
T_{n,p} Se_n(u,\bar{Q}_{n,p}) se_n(v,\bar{Q}_{n,p}) \nonumber
\end{array} 
\right\} \cos(k_{TM} z) ,
\end{eqnarray}
\begin{eqnarray}
&&E_u (u,v,z) =\frac{- k_{TM} }{\kappa_{1}^2 l(u,v)} \times\\
&& \times \left\{
\begin{array}{c}
D_{n,p} Ce_n'(u,Q_{n,p}) ce_n(v,Q_{n,p}) \\
T_{n,p} Se_n'(u,\bar{Q}_{n,p}) se_n(v,\bar{Q}_{n,p}) \nonumber
\end{array} 
\right\} \sin(k_{TM} z) ,
\end{eqnarray}
\begin{eqnarray}
&&E_v (u,v,z) =\frac{-k_{TM} }{\kappa_{1}^2 l(u,v)} \times\\
&& \times \left\{
\begin{array}{c}
D_{n,p} Ce_n(u,Q_{n,p}) ce'_n(v,Q_{n,p}) \\
T_{n,p} Se_n(u,\bar{Q}_{n,p}) se'_n(v,\bar{Q}_{n,p}) \nonumber
\end{array} 
\right\} \sin(k_{TM} z) .
\end{eqnarray}
From these, the components of the magnetic field are:
\begin{eqnarray}
H_z (u,v,z) = 0
\end{eqnarray}
\begin{eqnarray}
&&H_u (u,v,z) =\frac{-\varepsilon_0 \omega }{\kappa_{1}^2 l(u,v)} \times \\
&& \times \left\{
\begin{array}{c}
D_{n,p} Ce_n(u,Q_{n,p}) ce'_n(v,Q_{n,p}) \\
T_{n,p} Se_n(u,\bar{Q}_{n,p}) se'_n(v,\bar{Q}_{n,p}) \nonumber
\end{array} 
\right\} \cos(k_{TM} z) ,
\end{eqnarray}
\begin{eqnarray}
&&H_v (u,v,z) =\frac{\varepsilon_0 \omega }{\kappa_{1}^2 l(u,v)} \times \\
&& \times \left\{
\begin{array}{c}
D_{n,p} Ce'_n(u,Q_{n,p}) ce_n(v,Q_{n,p}) \\
T_{n,p} Se'_n(u,\bar{Q}_{n,p}) se_n(v,\bar{Q}_{n,p}) \nonumber
\end{array} 
\right\} \cos(k_{TM} z) .
\end{eqnarray}

Similar as before, the frequencies of the TM modes can be written in the form
\begin{eqnarray}
\omega_{n,p}^{TM}=\frac{1}{\mu_0 \varepsilon_0} \sqrt{\left(\frac{\pi m}{L}\right)^2 + 4\frac{Q(n,p)}{f^2}},
\label{eq:TMfreq}
\end{eqnarray}
where $Q(n,p) = Q_{n,p}$ and $ \bar{Q}_{n,p}$ for a specific mode.
\subsection{Comparison with experiment}
Experimental determination of the eigenmode frequencies was done using a copper cavity of height $L=28$ mm, and lengths of the major and minor axes $2a=21$ mm and $2b=13$ mm. This cavity was used for microwave absorption measurements (Refs.~\onlinecite{Pozek01,Pozek02,Pozek03,Pozek06,Grbic09,Grbic11,Peligrad1998,Peligrad2001}), for which the sample was usually placed in the center of the cavity ($u=0$, $z=L/2$).
To determine the frequency of a particular eigenmode for the current work  we used only the empty cavity. The ac signal was emitted from a small hole with an antenna at the top of the cavity. Another small hole was used to place a receiver antenna. Both antennae were connected to a pair of semi-rigid coaxial cables. To reduce the losses in the cavity walls, the cavity was cooled to 4.2 K in a bath of liquid helium. The field distribution was determined from measurements on anisotropic samples mentioned earlier.

\begin{table}[h]
	\begin{ruledtabular}
		\begin{tabular}{ c c c c } 
			
			mode & $\omega_{exp}$ (GHz) & $\omega_{CM}$ (GHz) & $\omega_{th}$ (GHz) \\
			\hline
			$_e$TE$_{111}$ & 9.443 & 9.700 & 9.717  \\
			
			$_e$TE$_{112}$ & 13.148 & 13.412 & 13.232 \\
			
			$_o$TE$_{111}$ & 13.903 & 12.661 & 13.770  \\
			
			$_e$TM$_{010}$ & 13.700 & 12.698 & 14.175 \\
			
			$_e$TE$_{211}$ & 15.132 & 15.566 & 15.749 \\
			
			$_e$TM$_{011}$ & 15.329 & 13.905 & 15.093  \\
			
			$_e$TE$_{113}$ & 17.557 & 17.958 & 17.593 \\
			
			$_e$TE$_{212}$ & 18.509 & 18.122 & 18.130 \\
			
			$_e$TM$_{012}$ & 17.768 & 16.694 & 17.563 \\
		\end{tabular}
	\end{ruledtabular}
	\caption{Comparison of the resonant microwave cavity eigenmodes for several lowest values. Frequencies are determined by experiment ($\omega_{exp}$), numerical analysis with a Comsol Multiphysics program package ($\omega_{CM}$), and analytical solution of the current calculation ($\omega_{th}$). Index $e$ marks that the orientation of the electric field is along the shorter axis of the ellipse, while $o$ indicates it is along the longer axis. }
	\label{tab:10mod}
\end{table}
The calculated modes of the cavity were compared to the experimental values, and numerical simulations, with the results shown in Table~\ref{tab:10mod}, where several of the lowest eigenmodes of the cavity are listed. The agreement between experiment and calculation is very good, given that the cavity walls are not perfect conductors, and that the cavity has additional small holes. 
\begin{figure}[h!]
		\centering
		\includegraphics[width=0.23 \textwidth]{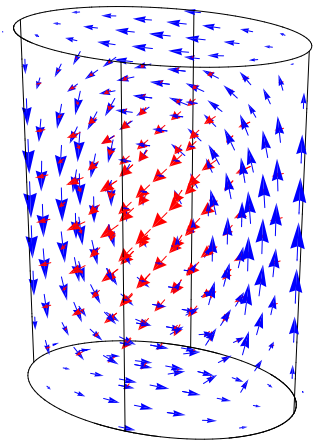}
	\caption{Calculated distribution of the magnetic (blue arrow) and electric (red arrows) field for the $_e$TE$_{111}$ eigenmode. The edges represent the boundary of the resonant cavity. \label{Fig111}}
\end{figure}

\begin{figure}[h!]

		\centering
		\includegraphics[width=0.23 \textwidth]{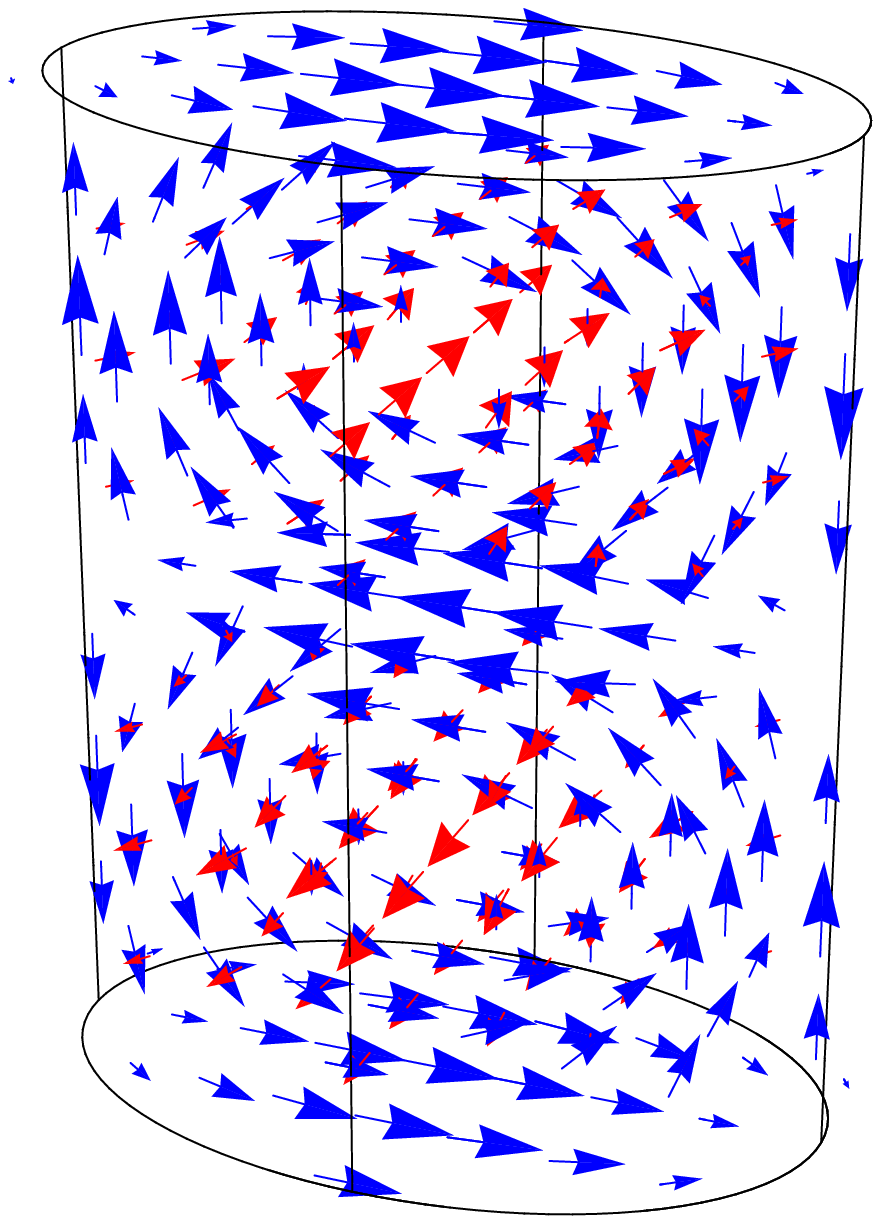}
	\caption{Calculated distribution of the magnetic (blue arrow) and electric (red arrows) field for the $_e$TE$_{112}$ eigenmode. The edges represent the boundary of the resonant cavity. \label{Fig112}}
\end{figure}

\begin{figure}[h!]
	\centering
	\includegraphics[width=0.25 \textwidth]{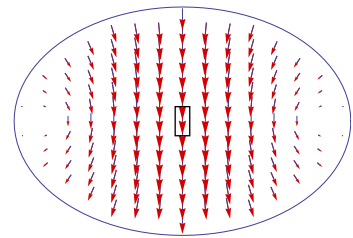}
	\caption{Two-dimensional electric field distribution at $L/2$ (sample is marked with a black rectangle) for the $_e TE_{111}$ mode. The size of the arrows is proportional to the intensity of the field. The elliptic edge shows the boundary of the resonant cavity.\label{figTE1112D}}
	\vspace{-10pt}
\end{figure}
Another advantage of the analytical solution is that we can easily check the electromagnetic field homogeneity at the sample position and for specific sample dimensions. For example, the eigenmodes we used the most were the $_e$TE$_{111}$ and the $_e$TE$_{112}$, for which we show the field distribution in Figs.~\ref{Fig111} and ~\ref{Fig112}, respectively. It has been previously shown\cite{Peligrad1998} for metallic and superconducting samples that as long as the sample volume is much smaller than the volume of the cavity, the field distribution is only slightly perturbed from the empty-cavity case. And as such we can study the homogeneity of the magnetic(electric) fields at the sample position using the eigenmodes of the cavity. 

\begin{figure}[t!]
		\centering
	\includegraphics[width=0.271\textwidth]{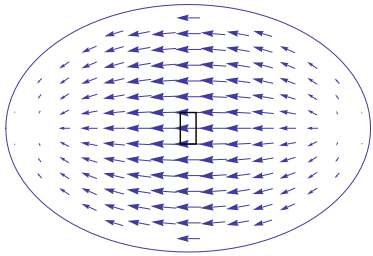}	
	\caption{Two-dimensional magnetic field distribution at $L/2$ (sample is marked with a black rectangle) for the $_e TE_{112}$ mode. The size of the arrows are proportional to the intensity of the field. The elliptic edge shows the boundary of the resonant cavity.\label{figTE1122D}}
\end{figure}

The typical size of the sample's cross-section is $1\times2$ mm$^2$. For the $_e$TE$_{111}$ and $_e$TE$_{112}$ eigenmodes the field distribution in the plane with the sample is shown in Figs.~\ref{figTE1112D} and ~\ref{figTE1122D}. Normally, the thickness of the sample is 1 mm or less, and the height of the cavity is its largest dimension - hence, the homogeneity in the $z$ direction is not questionable.

Homogeneity of the field on the sample site is important to check since one of the  advantages of the microwave absorption technique is that it can avoid problems of inhomogeneous current injection, which is a more difficult task for the dc transport where electrical contacts are used.

\section{Evolution of the eigenmodes}
We can now study how particular eigenmodes evolve from the circular cylindrical geometry. As we mentioned earlier, the advantage of the elliptical cavity is that it breaks the degeneracy of certain modes (e.g., $_o$TE$_{111}$ and $_e$TE$_{111}$ modes from Table~\ref{tab:10mod}) which allows larger control of the measurement setup, but it also enables the study of frequency dependent response for particular modes - for instance, if we want to study the sample in the electric field antinode, we can study the response for $_e$TE$_{111}$ and $_e$TE$_{113}$ modes if the sample is in the cavity center. Such type of measurement control is not that easy for a circular cylindrical cavity since different samples can perturb the modes enough so they cross in frequency. This can happen in the case of a ferroelectric or ferromagnetic samples, or a sample with large anisotropy of electric conductivity.
\begin{figure}[t!]
	\centering
	\includegraphics[width= \linewidth]{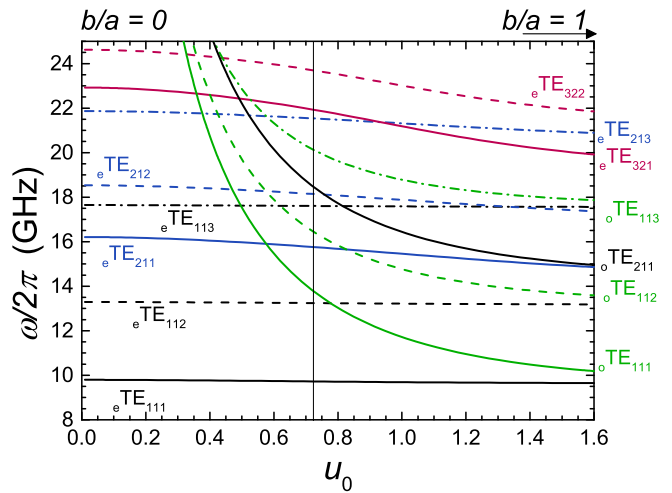}
	\caption{Dependence of TE modes frequency on the $u_0$ parameter from the circular $u_0 = 1.6$ ($b/a \approx 1$) to the elliptical $u_0 = 0$ ($b/a=0$) limit. The vertical line at $u_0=0.723$ marks the position of the microwave cavity used for experiments. \label{figTEmodovi}}
\end{figure}
To study how the eigenmodes evolve when the cross-section of the cavity changes from the circular to elliptic cylindrical geometry, we have calculated roots of the boundary conditions for a hypothetical cavity of the same height, albeit with different ratios of the major and minor axes. This allows a more deterministic approach in microwave cavity designs. The analysis was done by searching for the roots of the boundary conditions (\ref{eq:boundTE}), (\ref{eq:boundTE2}), (\ref{eq:boundTM}) and  (\ref{eq:boundTM2}) for different values of the parameter $u_0$. In the limit of $b/a \rightarrow 1$ the cavity has circular cross-section and parameter $u_0 \rightarrow \infty$, while in the limit of $b/a = 0$ the cavity has extremely elliptical cross-section and parameter $u_0 = 0$.
\begin{figure}[h!]
	\centering
	\includegraphics[width=0.45 \linewidth]{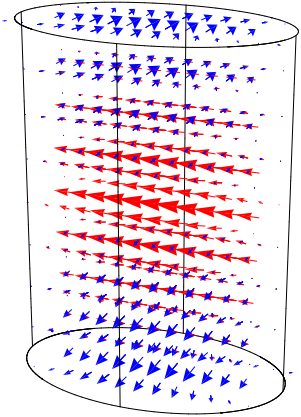}
	\caption{Calculated distribution of the magnetic (blue arrow) and electric (red arrows) field for the $_o$TE$_{111}$ eigenmode. \label{Figo111}}
\end{figure}\\
\indent The eigenmode frequencies are shown in figures~\ref{figTEmodovi} and ~\ref{figTMmodovi}, while the roots of the boundary conditions for $u_0 = 0.723$ (our measurement cavity) are listed in the Appendix~\ref{app:Roots}. In the graphs, the $u_0$ parameter varies up to $1.6$ that corresponds to $b/a = 0.92$, which is fairly close to the circular case. We can see how several modes (e.g. $_e$TE$_{111}$ and $_o$TE$_{111}$) become degenerate as the cross section of the cavity approaches the circular cylinder limit. From the field distribution in Fig.~\ref{Fig111} and~\ref{Figo111} we can easily visualize the problem mentioned earlier in the case of a circular cavity, when the two modes are degenerate. If the circular cavity is used to study e.g. cuprate superconductors, which have anisotropic $a$-, $b$- and $c$-axis conductivity, the field orientation will be determined by the sample properties as it will prefer one of the orientations. In this case the final field distribution cannot be easily predicted on the anisotropy value alone, since the geometry of the sample~\cite{Peligrad2001} also plays a role, but the edge-effects~\cite{UBCShape} as well. More complex situation arises when the temperature of the sample is changed, since the anisotropy of the sample changes as well. In the general case, the orientation can switch from the $_e$TE$_{111}$ to  $_o$TE$_{111}$ as the conductivity of the sample changes which makes it impossible to properly analyze the data. Hence, the measurement setup has to have a well defined electric
\begin{figure}[t!]
	\centering
	\includegraphics[width=\linewidth]{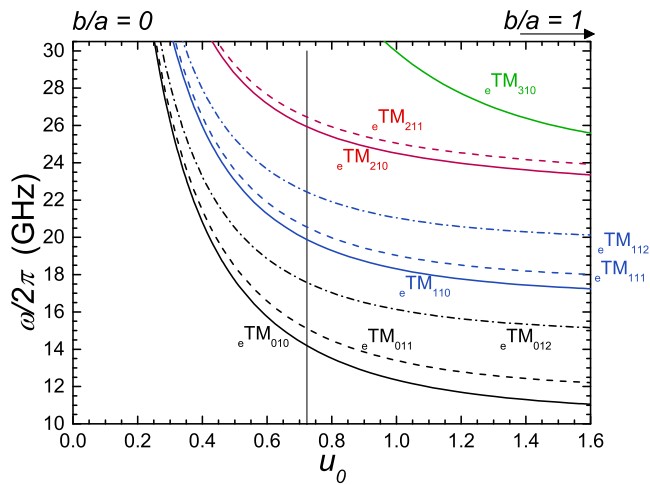}	
	\caption{Dependence of TM modes frequency on the $u_0$ parameter from the circular $u_0 = 1.6$ ($b/a \approx 1$) to the elliptical $u_0 = 0$ ($b/a=0$) limit. The vertical line at $u_0=0.723$ marks the position of the microwave cavity used for experiments. We show only the even modes since the odd ones are located at a higher frequency for this geometry.\label{figTMmodovi}}
\end{figure}
\begin{table}[t!]
	\begin{ruledtabular}
		\begin{tabular}{ c c c c c } 
			TE mode & $\omega^{TE}$ (GHz) && TM mode & $\omega^{TM}$ (GHz) \\
			\hline
			TE$_{111}$ & 9.940 && TM$_{010}$ & 10.935  \\
			
			TE$_{112}$ & 13.598 && TM$_{011}$ & 12.177 \\
			
			TE$_{211}$ & 14.886 && TM$_{012}$ & 15.309  \\
			
			TE$_{113}$ & 18.122 && TM$_{110}$ & 17.424 \\
			
			TE$_{212}$ & 17.541 && TM$_{111}$ & 18.229 \\
			
			TE$_{321}$ & 19.841 && TM$_{112}$ & 20.454  \\
			
			TE$_{213}$ & 21.241 && TM$_{210}$ & 23.353 \\
			
			TE$_{322}$ & 21.903 && TM$_{211}$ & 23.960			
		\end{tabular}
	\end{ruledtabular}
	\caption{Frequencies of several eigenmodes of a cylindrical cavity that correspond to our elliptical cavity in the limit $b/a=1$.}
	\label{tab:Cylmod}
	\vspace{-20pt}
\end{table}
(magnetic) field distribution.\\
\indent From Fig.~\ref{figTMmodovi} it can be seen that the frequencies of the TM modes vary more strongly with the shape of the boundary at $u_0$ than those of the TE modes. Expression (\ref{eq:TMfreq}) shows that the only parameters that influence the frequency dependence are the height of the cavity $L$, distance $f=\sqrt{a^2-b^2}$ and the value of the root $Q$ defined by the boundary conditions. Since we keep the height of the cavity constant and for $u_0\rightarrow0$ the value of $f\rightarrow a$, it follows that the shown frequency dependence is a direct consequence of the boundary conditions and the properties of the Mathieu functions. The physical interpretation of this trend is that the magnetic field of the TM modes, that is parallel to the elliptic base, requires shorter wavelengths to compensate for the higher curvature at the $v=0$ and $\pi$ points of the cavity in the limit $b/a \rightarrow 0$. This is not the case for the $_e$TE modes, which are defined by the direction of the electric field that connects the two semi-elliptic surfaces along the shorter axis. In the limit $b/a \rightarrow 0$ the area of these surfaces saturate in value, which allows for the frequency to remain practically constant.\\
\indent To verify that in the limit $u_0 = 1.6$ the system does behave as a circular cylinder (with 21 mm in diameter and height of 28 mm) and conclude this section, we have also calculated\cite{MIT11} the frequencies of the modes shown in Figs.~\ref{figTEmodovi} and~\ref{figTMmodovi} for the circular cavity and present them in Table~\ref{tab:Cylmod}. As it can be seen, the frequencies properly correspond to the modes of the elliptical cavity for $u_0 = 1.6$.
\section{Q-factor of certain eigenmodes}
In the development of the resonator cavity, one of the important quantities to monitor is the Q-factor - the measure of losses in the system. In general, there are many contributions to the Q-factor depending on the origin of the dissipation. We will calculate the contribution to the Q-factor arising from a specific mode in an empty cavity made out of lossy metallic material, which is also known as the \textit{unloaded} Q-factor. It is calculated by evaluating~\cite{comment}
\begin{equation}
Q= \frac{2}{\delta} \frac{\int_V \textbf{H}^2dV}{\int_S |\hat{n} \times \textbf{H}|^2dS},
\end{equation}
where $\delta$ is the skin-depth in the wall of the cavity, and the surface integral takes the tangential components of \textbf{H}, which is why it is represented as a vector product where $\hat{n}$ is the normal of the surface inside the cavity. We will take, as is usually done, that the finite conductivity of the wall does not influence the distribution of the fields previously calculated for a lossless cavity, but that the losses of the modes will be dominated by the skin-depth penetration of the fields into the walls of the cavity.

For the TE modes the integral in the numerator can be written as 
\begin{eqnarray}
&&\int_V \textbf{H}^2dV = \int\limits_0 ^{u_0} {\int\limits_0 ^{2\pi}} {\int\limits_0 ^{L}} l(u,v) ^2 \left( | H_z (u,v,z) | ^2 + \right. \nonumber \\
+ && \left. | H_u (u,v,z) |^2  + | H_v (u,v,z) | ^2 \right) du dv dz,
\end{eqnarray}
where $l(u,v)$ has been introduced earlier, and is also know as the scale factor for elliptical coordinates. The denominator can now be expanded into:
\begin{widetext}
\begin{eqnarray}
&&\frac{2}{\delta} \int |\hat{n} \times \textbf{H}|^2 dS  = \frac{2}{\delta} 
\left\{ \int\limits_0 ^{u_0} \int\limits_0 ^{2 \pi} l(u,v) ^2 (|H_u (u,v,0) |^2  +  | H_v (u,v,0) | ^2) dudv + \right.\nonumber \\
+ && \left. \int\limits_0 ^{u_0} \int\limits_0 ^{2 \pi} l(u,v) ^2 (|H_u (u,v,L) |^2  + | H_v (u,v,L) | ^2) dudv + \int\limits_0 ^{L} \int\limits_0 ^{2 \pi} l(u_0,v) (|H_z (u_0,v,z) |^2  + | H_v (u_0,v,z) | ^2)dzdv \right\}.
\end{eqnarray}
\end{widetext}
The first two parts of the last expression come from the contribution across the bases of the cylinder, while the last one from the side.

In both these integrals the $z$ dependence is simple and can be integrated-out, which simplifies our expressions. For the numerator it then follows:

\begin{eqnarray}
\int_V \textbf{H}^2dV = \frac{L}{2}&&\int\limits_0 ^{u_0} {\int\limits_0 ^{2\pi}} l(u,v) ^2 \left( | H_z (u,v) | ^2 + \right. \nonumber \\
+ && \left. | H_u (u,v) |^2  + | H_v (u,v) | ^2 \right) du dv,
\end{eqnarray}
and for the denominator:
\begin{eqnarray}
&&\frac{2}{ \delta} \int_S |\hat{n} \times \textbf{H}|^2 dS = \nonumber \\ &&\frac{2}{ \delta}
\left\{ 2 \int\limits_0 ^{u_0} \int\limits_0 ^{2 \pi} l(u,v) ^2 (|H_u (u,v) |^2  + | H_v (u,v) | ^2) dudv \right. + \nonumber \\ 
&&+ \left. \frac{L}{2} \int\limits_0 ^{2 \pi} l(u_0,v) (|H_z (u_0,v) |^2  + | H_v (u_0,v) | ^2)dv \right\},
\end{eqnarray}
where it is visible that the contributions at both bases are the same.

For the TM modes these expressions are even simpler since $H_z$ vanishes by definition and does not contribute to the integrals:
\begin{eqnarray}
\int_V \textbf{H}^2dV = \frac{L}{2}\int\limits_0 ^{u_0} {\int\limits_0 ^{2\pi}} &&l(u,v) ^2 \left( | H_u (u,v) | ^2 + \right. \nonumber \\
+ && \left. | H_v (u,v) | ^2 \right) du dv,
\end{eqnarray}
\begin{eqnarray}
&&\frac{2}{ \delta} \int_S |\hat{n} \times \textbf{H}|^2 dS = \nonumber \\ &&\frac{2}{ \delta}
\left\{ 2 \int\limits_0 ^{u_0} \int\limits_0 ^{2 \pi} l(u,v) ^2 (|H_u (u,v) |^2  + | H_v (u,v) | ^2) dudv \right. + \nonumber \\ 
&&+ \left. \frac{L}{2} \int\limits_0 ^{2 \pi} l(u_0,v)  | H_v (u_0,v) | ^2 dv \right\}.
\end{eqnarray}
Here, we immediately wrote the expressions with the $z$ dependence integrated out.

\begin{figure}[t!]
	\centering
	\includegraphics[width=0.49 \textwidth]{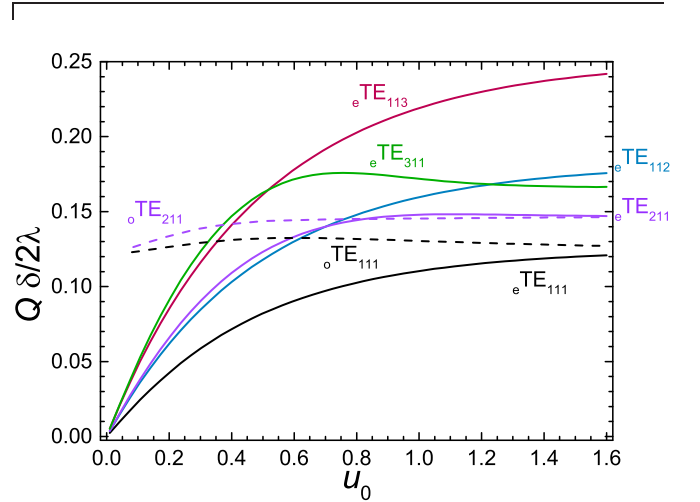}
	\caption{Q-factor for several TE modes calculated from the circular ($u_0\approx 1.6$) to the elliptical ($u_0 \rightarrow 0$) limit, as in the previous figures. Some of the characteristics of these modes are shown in previous figures. Note how the Q-factor of $_o$TE$_{111}$ and $_e$TE$_{111}$ modes converge in the $b/a\rightarrow 1$ limit ($u_0 \approx 1.6$). \label{figQTE}}
\end{figure}

\begin{figure}[h!]
	\centering
	\includegraphics[width=0.49 \textwidth]{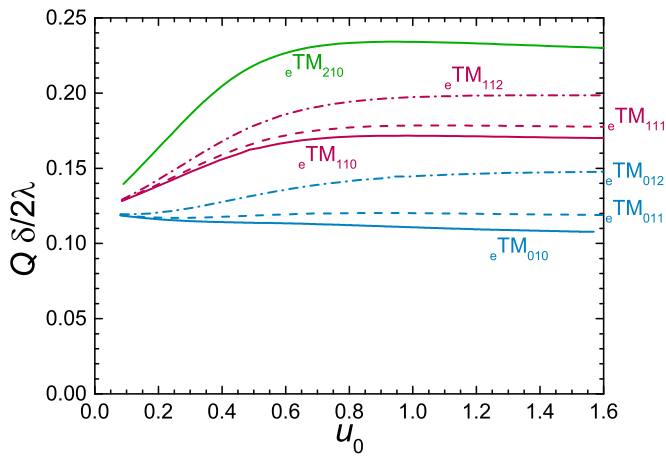}
	\caption{Q-factor for several TM modes calculated from the circular ($u_0\approx 1.6$) to the elliptical ($u_0\rightarrow 0$) limit, as in the previous figures.  \label{figQTM}}
\end{figure}

To analyze how the Q-factors evolve together with the eigenmodes when the $b/a$ ratio changes, we need to calculate the above integrals, despite attempts we were not able to find an analytical solution for some of them. In particular, the integrals of $\int_S |\hat{n} \times \textbf{H}|^2dS$ that contain  $l(u,v)^2$ can be analytically solved, but the solution does not simplify much the calculation procedure (Appendix~\ref{app:analyticI}). On the other hand, the integrals containing only $l(u,v)$ are more difficult. Therefore the Q-factors were evaluated using numerical integration procedures. The results for several modes of interest are shown in figures~\ref{figQTE} and \ref{figQTM}. We present the dimensionless quantity $Q \delta /2\lambda$, as is usually done, since it depends only on the mode and the cavity shape.
\\	
\indent When these dependences are qualitatively compared to those~\cite{Poole,MIT11} for a circular cylinder and a rectangular box, it can be seen that the curves shown in Figs.~\ref{figQTE} and ~\ref{figQTM} are a combination of the two. By this we mean that in the limit $u_0 \rightarrow 0$ our system resembles a thin rectangular cavity, while in the other a circular cylinder. Therefore, for $u_0 \rightarrow 0$ the energy in the cavity volume diminishes and the dissipation remains finite, which causes Q$\rightarrow 0$ for the $_e$TE modes, while for $u_0$ large the Q-factor remains finite. Similarly, for $u_0 \rightarrow 0$ the value of $Q \delta /2\lambda$ for the $_o$TE and the TM modes saturates, since the divergence of the eigenfrequency occurs equally rapid as reduction of energy in the cavity volume. 
\section{Conclusion}
We have calculated and presented an analytical solution for TE and TM eigenmode frequencies and field distributions of an elliptical cylindrical cavity. We have compared our results for the case of a microwave absorption measurement technique, where we find good matching between calculated, measured and numerically simulated frequency values.

We have also shown several eigenmode field distributions that can be easily connected to the related modes of the circular cylinder with broken degeneracy. By following their evolution as the ellipticity of the cross-section is reduced we have shown that the modes degeneracy reduces and the eigenmode frequencies eventually converge to their circular cross-section limit.

Furthermore, we have shown how our solutions can be used for calculation of the unloaded Q-factors and how these values also depend on the shape of the elliptical cross section.

\begin{acknowledgments}
We acknowledge the support of Croatian Science Foundation (HRZZ) under the project 2729, the Unity Through Knowledge Fund (UKF Grant No. 20/15) and the support of project CeNIKS co-financed by the Croatian Government and the European Union through the European Regional Development Fund - Competitiveness and Cohesion Operational Programme (Grant No. KK.01.1.1.02.0013). We also acknowledge the advice and discussions with A.~Dul\v{c}i\'{c}, M.~Basleti\'{c} and M.~Po\v zek. The author acknowledges Ted Ersek for the help and discussions on calculating the roots of the Mathieu functions.
\end{acknowledgments}
\appendix
\section{Separation of variables $u,v$}
\label{app:Separ}
The equation of the general form
\begin{eqnarray}
\frac{\partial^2 E_z}{\partial u^2}+\frac{\partial^2 E_z}{\partial v^2}&&+\frac{1}{2} {\kappa_{1}}^2 f^2  \nonumber \\
&&\times (\cosh(2 u)-\cos(2 v)) E_z = 0,
\end{eqnarray}
can be simplified by letting $E_z (u,v) = U(u) V(v)$, which leads to
\begin{eqnarray}
U''V&&+UV'' \nonumber \\
&&+\frac{1}{2} {\kappa_{1}}^2 f^2 (\cosh(2 u)-\cos(2 v)) UV = 0,
\end{eqnarray}
and further
\begin{eqnarray}
\frac{U''}{U}+\frac{V''}{V}+{\kappa_{1}}^2 f^2 (\cosh^2(u)-\cos^2(v)) = 0.
\end{eqnarray}
Clearly, this equation can be further separated by introducing the separation constant $\xi$:
\begin{eqnarray}
\frac{U''}{U}+{\kappa_{1}}^2 f^2 &&\cosh^2(u) =\nonumber \\
&&= -\frac{V''}{V}+{\kappa_{1}}^2 f^2\cos^2(v)= \xi.
\end{eqnarray}
This leads us to the form of the Mathieu and modified Mathieu equation:
\begin{eqnarray}
V''+(\xi-G^2\cos^2(v))V= 0, \\
U''-(\xi- G^2 \cosh^2(u) )U=0,
\end{eqnarray}
respectively, where we substituted $G^2 = {\kappa_{1}}^2 f^2$. The values of $\xi$ are the so-called \textit{characteristic number} $a(n,q)$  for the functions $ce_n$ and $Ce_n$, and $b(n,q)$ for the functions $se_n$ and $Se_n$.
\section{Roots of the boundary conditions $q_{n,p}, \bar{q}_{n,p}, Q_{n,p}$ and $\bar{Q}_{n,p}$}
\label{app:Roots}
Here we list all the roots of the modified Mathieu functions and its first derivative for $q \leq 100$ and $u_0 = 0.7235$ which matches the dimensions of our resonant cavity:
	\begin{table}[h]
		\begin{ruledtabular}
		\caption{TE modes $Ce'_{n,p} (u_0, q) = 0$}
		\begin{tabular}{ c c c c c c } 
			$n$ & $q_{n,1}$& $q_{n,2}$& $q_{n,3}$& $q_{n,4}$& $q_{n,5}$ \\
			\hline
			0 & 0.0 & 4.76358 & 17.5302 & 38.2335 & 66.8774 \\			
			1 & 0.537555 & 6.96811 & 21.4144 & 43.8068 & 74.139\\			
			2 & 1.76029 & 9.79541 & 25.8556 & 49.923 & 81.9364 \\			
			3 & 3.60884 & 13.302 &  30.8798& 56.6004 & 90.2837 \\			
			4 & 6.03824 & 17.5238 & 36.5105 & 63.8558 & 99.1944 \\
		\end{tabular}
	
		\caption{TE modes $Se_{n,p} ' (u_0, q) = 0$}
		\begin{tabular}{ c c c c c c }
			$n$ & $\bar{q}_{n,1}$& $\bar{q}_{n,2}$& $\bar{q}_{n,3}$& $\bar{q}_{n,4}$& $\bar{q}_{n,5}$ \\
			\hline
			1 & 1.29523 & 10.155 & 26.8894 & 51.5628 & 84.1774 \\	
			2 & 2.49731 & 13.1934 & 31.6183 & 57.9803 & 92.2831 \\
			3 & 4.22963 & 16.8007 & 36.8957 & 64.9366 & - \\
			4 & 6.50922 & 21.0059 & 42.743 & 72.4478 & - \\
			5 & 9.33416 & 25.8297 & 49.1795 & 80.5288 & - \\
		\end{tabular}
	\end{ruledtabular}
	\end{table}
%
\begin{table}[h]
\begin{ruledtabular}
	\caption{TM modes $Ce_{n,p} (u_0, q) = 0$}
	\begin{tabular}{c c c c c c} 
		$n$ & $Q_{n,1}$& $Q_{n,2}$& $Q_{n,3}$& $Q_{n,4}$& $Q_{n,5}$ \\
		\hline
		0 & 1.59922 & 10.3356 & 27.0588 & 51.7276 & 84.3399 \\
		1 & 3.14128 & 13.4264 & 31.8181 & 58.1666 & 92.462\\
		2 & 5.34388 & 17.0901 & 37.1269 & 65.145 & - \\
		3 & 8.2234 & 21.3596 & 43.0059 & 72.6787 & - \\
		4 & 11.7793 & 26.2669 & 49.4738 & 80.7823 & - \\	
	\end{tabular}

	\caption{TM modes $Se_{n,p} (u_0, q) = 0$}
	\begin{tabular}{c c c c c c}
		$n$ & $\bar{Q}_{n,1}$& $\bar{Q}_{n,2}$& $\bar{Q}_{n,3}$& $\bar{Q}_{n,4}$& $\bar{Q}_{n,5}$ \\
		\hline
		1 & 4.95508 & 17.7036 & 38.4002 & 67.0409 & -\\
		2 & 7.21575 & 21.6261 & 43.9987 & 74.3211 & - \\
		3 & 10.0515 & 26.1064 & 50.1408 & 82.1377 & - \\
		4 & 13.4815 & 31.1687 & 56.8446 & 90.5046 & - \\
		5 & 17.5135 & 36.8336 & 64.1264 & 99.4351 & - \\	
	\end{tabular}
\end{ruledtabular}
\end{table}
\section{Integrals that can be solved analytically}
\label{app:analyticI}
In the integrals with $l(u,v)^2$ this expression is canceled by its inverse present in magnetic field components $(H_u, H_v, H_z)$, and these integrals can than be sorted using the identities~\cite{McLachlan}:
\begin{eqnarray}
&&\int_0 ^{2\pi} ce_n^2(v,q)dv = \int_0 ^{2\pi} se_n^2(v,q)dv = \pi, \\
&&\int_0 ^{2\pi} {ce_n '}^2 (v,q)dv = \vartheta_n \pi, \\
&&\int_0 ^{2\pi} {se_n'}^2 (v,q)dv = \overline{\omega}_n \pi,
\end{eqnarray}
where $\vartheta_n = a(n,q) - 2 q~\Theta_n$ and $\overline{\omega}_n = b(n,q) - 2 q~\Psi_n$. Functions $\Theta_n$ and $\Psi_n$ match other identities useful for other integrals and can be evaluated by the following sums:
\begin{eqnarray}
\Theta_{2n} &&= \frac{1}{\pi} \int _0 ^{2\pi}ce_{2n} ^2 (v) \cos (2v)dv \\ 
&&= A_0 ^{(2n)} A_2 ^{(2n)} + \sum_{r=0} ^\infty A_{2r} ^{(2n)} A_{2r+2} ^{(2n)}, \\
\Theta_{2n+1} &&= \frac{1}{\pi} \int _0 ^{2\pi}ce_{2n+1} ^2 (v) \cos (2v)dv\\
&&=\frac{1}{2} \left[ A_1 ^{(2n+1)} \right]^2 + \sum_{r=0} ^\infty A_{2r+1} ^{(2n+1)} A_{2r+3} ^{(2n+1)},
\end{eqnarray}
\begin{eqnarray}
\Psi_{2n+2} &&=\frac{1}{\pi} \int _0 ^{2\pi}se_{2n+2} ^2 (v) \cos (2v)dv,\\
&&= \sum_{r=0} ^\infty B_{2r+2} ^{(2n+2)} B_{2r+4} ^{(2n+2)}, \\
\Psi_{2n+1}  &&= \frac{1}{\pi} \int _0 ^{2\pi}se_{2n+1} ^2 (v) \cos (2v)dv, \\
&&=-\frac{1}{2} \left[ B_1 ^{(2n+1)} \right]^2 + \sum_{r=0} ^\infty B_{2r+1} ^{(2n+1)} B_{2r+3} ^{(2n+1)}.
\end{eqnarray}
The coefficients $A_{r} ^n$ and $B_{r} ^n$ can be calculated by a recursion relation\cite{McLachlan}.

In addition, for integral containing $Ce_n(u,q)$ and $Se_n(u,q)$ the identity
\begin{equation}
\int_0 ^{u_0} y^2(u)\cosh(2u)du = \frac{1}{2} \left[\frac{\partial y}{\partial u} \frac{\partial y}{\partial q} - y \frac{\partial}{\partial u} \left( \frac{\partial y}{\partial q} \right)\right]_0 ^{u_0},
\end{equation}
can be used.

\begin{thebibliography}{21}
\bibitem [{com()}]{comment}%
	\BibitemOpen
	\href@noop {} {}\bibinfo {note} {Calculation of these integrals in elliptic
		coordinates includes a 'nontrivial' {J}acobian determinant. {A}s such,
		surface and volume elements need to be introduced with appropriate scaling
		factors, which was in our impression not taken into account in previous
		work\cite{Chu,Kinzer,Higgins}}\BibitemShut {NoStop}%
\end{thebibliography}
%

\end{document}